\documentclass[twocolumn,english,aps,prd,graphics,floatfix,a4paper,superscriptaddress]{revtex4-1} 
\usepackage{amsmath, amssymb}
\usepackage{makecell}
\usepackage{multirow}
\usepackage{CJK}
\usepackage{graphicx}
\usepackage{mathrsfs}
\usepackage{bm}
\usepackage{amsmath}
\usepackage{dcolumn}
\usepackage{epstopdf}
\usepackage{dsfont}
\usepackage{amssymb}
\usepackage{tabularx}
\usepackage{array}
\usepackage{float}
\usepackage{color}
\usepackage{epstopdf}
\usepackage{mathrsfs}

\usepackage[colorlinks, linkcolor=blue,anchorcolor=blue,citecolor=blue,urlcolor=blue]{hyperref}
\usepackage{extarrows}
\usepackage{threeparttable}
\begin{document}
	
\title{Chiral topological metals with multiple types of quasiparticle fermions and large spin Hall effect in the SrGePt family materials}

\author{Yi Shen}\email{These authors contributed equally to this work.}
	\affiliation{School of Physics and Electronics, Hunan Normal University, Key Laboratory for
Matter Microstructure and Function of Hunan
Province, Key Laboratory of Low-Dimensional Quantum Structures and Quantum Control of Ministry of Education, Changsha 410081, China}

\author{Yahui Jin}\email{These authors contributed equally to this work.}
\affiliation{School of Physics and Electronics, Hunan Normal University, Key Laboratory for
Matter Microstructure and Function of Hunan
Province, Key Laboratory of Low-Dimensional Quantum Structures and Quantum Control of Ministry of Education, Changsha 410081, China}	
\affiliation{Department of Physics, Changji University, Xinjiang, Changji 831100, China}

\author{Yongheng Ge}
	\affiliation{School of Physics and Electronics, Hunan Normal University, Key Laboratory for
		Matter Microstructure and Function of Hunan
Province, Key Laboratory of Low-Dimensional Quantum Structures and Quantum Control of Ministry of Education, Changsha 410081, China}	

\author{Mingxing Chen}
\affiliation{School of Physics and Electronics, Hunan Normal University, Key Laboratory for
	Matter Microstructure and Function of Hunan
Province, Key Laboratory of Low-Dimensional Quantum Structures and Quantum Control of Ministry of Education, Changsha 410081, China}

\author{Ziming Zhu}\email{zimingzhu@hunnu.edu.cn}
\affiliation{School of Physics and Electronics, Hunan Normal University, Key Laboratory for
	Matter Microstructure and Function of Hunan
Province, Key Laboratory of Low-Dimensional Quantum Structures and Quantum Control of Ministry of Education, Changsha 410081, China}


\begin{abstract}
We present a prediction of chiral topological metals with several classes of unconventional quasiparticle fermions in a family of SrGePt-type materials in terms of first-principles calculations. In these materials, fourfold spin-3/2 Rarita-Schwinger-Weyl (RSW) fermion, sixfold excitation, and Weyl fermions coexist around the Fermi level as spin-orbit coupling is considered, and the Chern number for the first two kinds of fermions is the maximal value four. We found that large Fermi arcs from spin-3/2 RSW fermion emerge on the (010)-surface, spanning the whole surface Brillouin zone. Moreover, there exist Fermi arcs originating from Weyl points, which further overlap with trivial bulk bands. In addition, we revealed that the large spin Hall conductivity can be obtained, which attributed to the remarkable spin Berry curvature around the degenerate nodes and band-splitting induced by spin-orbit coupling. Our findings indicate that the SrGePt family of compounds provide an excellent platform for studying on topological electronic states and the intrinsic spin Hall effect.

\end{abstract}

\maketitle
\section{Introduction}
Topological metals (TMs) or semimetals have attracted a large variety of research interests~\cite{wanWeyl,wengprx,huang2015weyl,soluyanov2015type,PhysRevLett.108.140405,RevModPhys.88.035005,RevModPhys.90.015001,RevModPhys.93.025002,ge2022ferromagnetic} for the past decade owing to the exotic electronic properties such as unique magneto-transport and bulk photogalvanic response~\cite{BurkovPRB,xiong2015evidence,PhysRevX.5.031023,de2017quantized,changprl,tangprl,ni2021giant}. The appearance of band degeneracies like Dirac points or Weyl points in TMs can be stabilized by symmetries or band topology, around which the quasiparticle excitations resemble the analogy of Dirac or Weyl fermions in high energy physics~\cite{RevModPhys.88.035005,RevModPhys.90.015001,RevModPhys.93.025002}. Compared with the elementary particles constrained by the poincar$\grave{e}$ symmetry, the respective crystal space group symmetries for the band-touching degenerate point in solids is much smaller, rendering the possibility for realizing unconventional particles beyond the standard model paradigms~\cite{bradlyn2016beyond,zhu2016triple,PhysRevLett.116.186402}. In particular, the unconventional fermions with Chern number larger than usual Weyl fermions could emerge in chiral crystals which lack inversion symmetry and mirror symmetry~\cite{changprl,tangprl}. Examples include threefold spin-1 excitations, double Weyl fermions, fourfold spin-3/2 RSW fermions and sixfold excitations, which feature the multiple Fermi-arcs associated with the number of topological charge.

So far, various works have focused on the predictions of TMs with multiple-types of unconventional fermions in the nonmagnetic system, and some of them are confirmed by ARPES observations~\cite{rao2019observation,sanchez2019topological,PhysRevLett.122.076402,schroter2019chiral,schroter2020observation}. For instance, CoSi-family materials are proposed to possess four- and six-fold band crossings at the time-reversal invariant momenta $\Gamma$ and R enforced by nonsymmorphic symmetries, exhibiting the long Fermi arcs spanning the surface Brillouin zone  (BZ) ~\cite{rao2019observation,sanchez2019topological,PhysRevLett.122.076402}. However, the degeneracy of nodal-points at the $\Gamma$ and R points in CoSi and RhSi are experimentally found to be three- and four-fold, since spin-orbit coupling (SOC) is rather weak such that the band-splitting can not be distinguished~\cite{rao2019observation,sanchez2019topological,PhysRevLett.122.076402}. Furthermore, two other experimental works have reported that AlPt~\cite{schroter2019chiral} and PdGa~\cite{schroter2020observation} exhibit multi-fold degenerate points with four of the Chern number, and their locations partially stay away from the Fermi level ($>$0.5 eV). Specially, the band-splitting of PdGa~\cite{schroter2020observation} induced by the substantial SOC is clearly identified by ARPES.
Thus, it is highly urgent to search for the new chiral TMs with large SOC as well as the energy positions of nodal-points closer to the Fermi level.

On the other hand, the electronic bands around the degenerate or quasi-degenerate points/lines in the TSMs can be regarded as the source of large spin Berry curvature (SBC), resulting in the large spin Hall effect (SHE). For example, it has been theoretically predicated that Weyl semimetal TaAs exhibits the large spin Hall conductivity (SHC) owing to the relatively strong SOC and Weyl points~\cite{sun2016strong}. Furthermore, several studies have reported that the large SHC can be achieved such as metallic rutile oxides~\cite{sun2017dirac}, ZrXY(X = Si, Ge; Y = S, Se, Te)~\cite{yen2020tunable} and so on, which result from the notable SBC around the gapped nodal-lines or points caused by SOC near the Fermi level. Therefore, it is very natural to further ask if chiral topological metals in the SrGePt family show the strong SHE due to the presence of chiral nodal-points and SOC-induced band-splitting.

In this work, based on the first-principles calculations, we systematically investigate on topological electronic states and SHC of the SrGePt family materials.  Take SrGePt as a typical example, we find it shows the coexistence of chiral spin-3/2 RSW fermion and six-fold excitation (two copies of spin-1 fermions) as well as Weyl points around the Fermi level, featuring nontrivial surface states for the (010) surface. Due to the prominent contribution of SBC around the nodal points and SOC-induced band-splitting, it has been demonstrated that the magnitude of the intrinsic SHC can reach up to 365 $(\hbar/e)(\Omega \cdot cm)^{-1}$ near the Fermi level. Our findings provide the compelling platform to study the interplay between the unconventional fermionic quasiparticles and spin transport in the condensed matter system.

\section{Computational Method}
Our first-principles calculations were based on  the density functional theory (DFT), using a plane-wave basis set and projector augmented wave method~\cite{PhysRevB.50.17953}, as implemented in the Vienna \emph{ab} \emph{initio} simulation package (VASP)~\cite{PhysRevB.54.11169,kresse1999ultrasoft}. The generalized gradient approximation (GGA) parameterized by Perdew, Burke, and Ernzerhof (PBE) was adopted for the exchange-correlation functional~\cite{PhysRevLett.77.3865}. The energy cutoff was set to 360 eV, and a $15\times15\times15$ Monkhorst-Pack $k$ mesh was used for the BZ sampling. The atomic positions were fully optimized until the residual forces were less than $10^{-3}$ eV/{\AA}. The convergence criterion for the total energy was set to be $10^{-8}$ eV. Band structure of SrGePt with SOC is further checked by the modified Becke-Johnson (mBJ) potential~\cite{mBJPRL}, which is consistent with the result using PBE potential.
To calculate the surface states and the intrinsic SHC, the tight-binding Hamiltonian (dubbed as Wannier-TB) was constructed by projecting the Bloch states onto \textit{d} orbit of Sr, \textit{p} orbit of Ge and \textit{d} orbit of Pt without performing the iterative spread minimization using the WANNIER90 package ~\cite{mostofi2014updated}. The outer energy window for the disentanglement is chosen to be from -1.0 eV to 13.0 eV, and the frozen energy is from -1.0 eV to 7.0 eV. The surface states were further investigated by using the iterative Green's function method~\cite{sancho1985highly} as implemented in the WannierTools package~\cite{wu2017wanniertools}.

Based on the Kubo formula, the intrinsic SHC can be evaluated by performing the integration of SBC over the whole BZ for the occupied bands~\cite{guo2005ab,guo2008intrinsic}. The intrinsic SHC at the clean limit can be expressed as:
\begin{equation}
\sigma_{ij}^{k}=e{\hbar}\sum_{n}\int_{BZ}\frac{d\mathbf{k}}{(2\pi)^3}f_{n\mathbf{k}}\Omega_{n,ij}^{s,k}(\mathbf{k})
\label{she}
\end{equation}
where $\Omega_{n,ij}^{s,k}(\mathbf{k})$ is the SBC for the $n$th band at $\mathbf{k}$
\begin{equation}
\Omega_{n,ij}^{s,k}(\mathbf{k})=-\sum_{n^{\prime}\neq{n}}\frac{2\mathrm{Im}[\langle{n}\mathbf{k}|\hat{J}_{i}^{k}|{n^{\prime}}\mathbf{k}\rangle\langle{n^{\prime}}\mathbf{k}|\hat{\upsilon}_{j}|n\mathbf{k}\rangle]}{(\epsilon_{n\mathbf{k}}-\epsilon_{{n^{\prime}}\mathbf{k}})^2}
\label{sbc}
\end{equation}
where the spin current operator $\hat{J}_{i}^{k}$=$\frac{1}{2}\left\{\hat{\upsilon}_{i},\hat{s}_{k}\right\}$, with the spin operator being $\hat{s}$=$\frac{\hbar}{2}\hat{\sigma}$ and velocity operator being $\hat{\upsilon}_{i}$. Indices $i$ and $j$ denote Cartesian directions, $k$ denotes the direction of spin, and $i$, $j$, $k$=$x$, $y$, $z$. $f_{n\mathbf{k}}$ is the Fermi-Dirac distribution function. A dense grid of 200$\times$200$\times$200 was adopted for the integral of SBC values in the BZ.

\section{Crystal structure and symmetries}
\begin{figure}[!htp]
	{\includegraphics[clip,width=8.2cm]{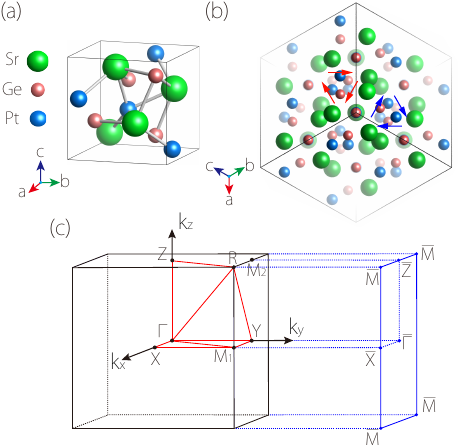}}
	\caption{\label{fig1}
(a) The perspective view in a unit cell and (b) top view with the supercell of $2\times2\times2$ along the (111)-axis. (c) Bulk and the projected surface BZ for the (010) surface in SrGePt. Here time-reversal invariant momenta are marked by the black and blue points. The red (blue) arrows for Ge (Pt) atoms in (b) indicate that the chirality of crystal structure is left-handed.}

\end{figure}

SrGePt-family materials include other six members: SrSiPd, BaSiPd, CaSiPt, SrSiPt, BaSiPt and BaGePt, which have been successfully synthesized by Evers et al. in 1992~\cite{evers1992ternare}. All of them share the same kind of cubic structure, with space group P2$_1$3 (No. 198). Since the similarity of physical properties among them, we mainly focus on SrGePt as a prototype for our discussions. SrGePt contains twelve atoms (i.e., four formula units) in a primitive unit cell, in which each Sr is bonded by two nearest neighboring (NN) Pt and two NN Ge as shown in Figs.~\ref{fig1} (a) and (b). The bulk BZ along with surface BZ are shown in Fig.~\ref{fig1} (c). In a unit cell, all atoms occupy at the Wyckoff positions of 4\emph{a}. The simulated lattice constants with $a = 6.692$ {\AA} were adopted in our calculations (see more in Appendix~\ref{simu_con}).

Lattice symmetries in SrGePt have chiral and non-symmorphic features, which contain three generators: threefold rotation symmetry along (111)-axis (${C_{3}^{111}}$) and two twofold screw rotation symmetries ${S}_{2x}=\{C_{2x}|\frac{1}{2}\frac{1}{2}0\}$ and ${S}_{2y}=\{C_{2y}|0\frac{1}{2}\frac{1}{2}\}$. Combining three of them, the total symmetry operators of SrGePt include three twofold and four threefold (screw) rotations. Note that fourfold rotation symmetry is broken in SrGePt system, although such lattice belongs to cubic system.
The studied crystal structures of SrGePt-family materials here belong to the left-handed crystal. Taking SrGePt as an example, the projected density of states (PDOS) shows that the low-energy bands near the Fermi level are dominated by Ge-\emph{p} and Pt-\emph{d} orbits as shown in Fig.~\ref{fig2}(a). Thus, we can define the chirality of SrGePt crystal structure according to Pt and Ge as reported in the Ref.~\cite{hsieh2022helicity}. As shown in Fig.~\ref{fig1}(b), the chirality of SrGePt is indeed left-handed, since both the Pt and Ge atoms have the left-handed chirality.

\section{Electronic structures with SOC}
\begin{figure}[!htp]
	{\includegraphics[clip,width=8.2cm]{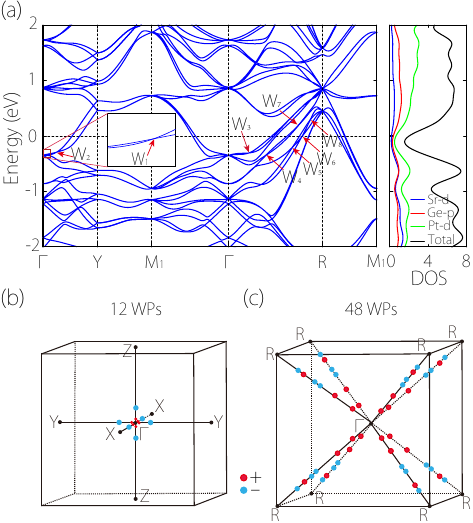}}
	\caption{\label{fig2} (a) Calculated band structure along the high symmetry directions and the total density of state along with PDOS for SrGePt in the presence of SOC. Distributions of Weyl points (WPs) along (b) $\Gamma$-X, $\Gamma$-Y and $\Gamma$-Z directions, as well as (c) $\Gamma$-R directions. Here red and wathet blue points indicate Weyl points with topological charges of +1 and -1, respectively.}
\end{figure}

In the absence of SOC, band structures of the SrGePt-family materials feature Weyl points along the $\Gamma$-R direction, quadruple Weyl node with $\mathcal{C}$=4 (QW) at the $\Gamma$ point and spin-1 Weyl node with $\mathcal{C}$=2 (SW) at the R point~\cite{PhysRevB.102.125148} (see Appendix~\ref{wosoc}). Since any compound for the SrGePt-family materials contains the heavy atom(s), we mainly focus on the calculations with SOC included in this work. Bulk band structure of SrGePt in the presence of SOC are illustrated in Fig.~\ref{fig2} (a). One can observe that SrGePt hosts metallic feature, with electron and hole pockets located at the $\Gamma$ and R points.  Band splitting occurs at the generic momenta except for the time-reversal invariant ones, since the inversion symmetry is broken. Intriguingly, due to the chiral and nonsymmorphic symmetries in SrGePt, unconventional quasiparticle excitations with multifold degeneracy such as spin-3/2 RSW fermions and time reversal doubling of spin-1 fermions are allowed as reported in Refs.~\cite{bradlyn2016beyond,tangprl,changprl}. Indeed, a fourfold degenerate point appears at the center of BZ ($\Gamma$), carrying topological charge of +4; while six-fold nodal-point arises at the boundary of BZ (R), and its topological charge is -4 [see Fig.~\ref{fig2} (a)]. It should be mentioned that such multifold nodal fermions beyond Dirac and Weyl fermions carry large topological charge, differentiating from the cases for triple point and cubic Dirac point~\cite{zhu2016triple,PhysRevLett.116.186402}, for which the chirality is ill-defined. Moreover, these nodal points with large Chern number have the energy offset about 0.8 eV thanks to the lack of mirror symmetry, which can lead to the occurrence of chiral photogalvanic transport~\cite{de2017quantized,ni2021giant}.
For the bands between fourfold and sixfold degenerate points, there are two Weyl points located at $\Gamma$-Y line  (W$_{1}$ and W$_{2}$) protected by twofold rotation and six Weyl points (W$_{3}$$\sim$W$_{8}$) at $\Gamma$-R line protected by threefold rotation, as shown in Fig.~\ref{fig2} (b) and (c).
Here band structure at $\Gamma$-Y is actually the same as those at $\Gamma$-X and $\Gamma$-Z. Thus, the total number of Weyl points are 60 along all high symmetry lines and their dispersions belong to type-II. Topological chiralities of Weyl fermions are indicated by red and wathet blue dots [see Figs.~\ref{fig2} (b) and (c)], and their sum is zero. In addition, we show the distributions of Berry curvature around W$_2$ and W$_3$, which indeed acts as the sink and source associated with the Chern number of -1 and +1 (see Appendix~\ref{BC_WP}). Band structures for other materials of the SrGePt-family exhibit the same topological features (see Appendix~\ref{wsoc}).

\begin{figure}[!htp]
	{\includegraphics[clip,width=8.2cm]{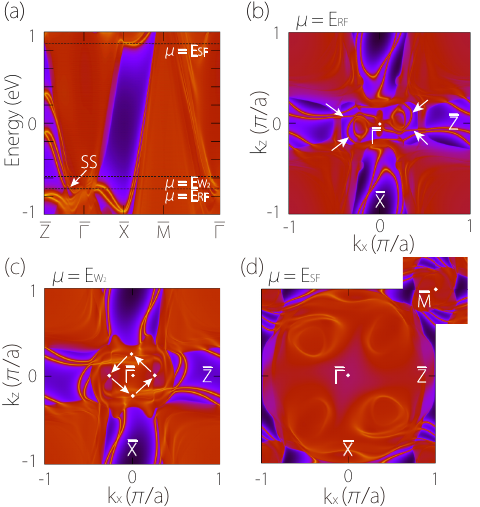}}
	\caption{\label{fig3} (a) Surface state of SrGePt for the (010) surface. (b)-(d) Constant energy slices of the surface spectrum for different energy. Here the contours cross the energy of spin-3/2 RSW fermion ($\rm E_{RF}$), Weyl point W$_2$ ($\rm E_{W_{2}}$)  and six-fold degenerte fermion ($\rm E_{SF} $), respectively. The white arrows in (b) indicate Fermi arcs relevant with spin-3/2 RSW fermion. The white arrows in (c) indicate the projected Weyl points W$_2$. Here Fermi energy is set to be zero.}
\end{figure}

\begin{table*}[t!]
\renewcommand\arraystretch{1.4}
  \caption {{Comparisons of SHC including $\sigma_{xy}^z$ and $\sigma_{xz}^y$ at the Fermi energy ($\rm E_{F}$) among all seven compounds of SrGePt family materials, ZrSiS, RhGe, TaAs and Pt. Here SHC have unit of $(\hbar/e)(\Omega \cdot cm)^{-1}$.}}
\begin{threeparttable}
\begin{tabular*}{\hsize}{@{}@{\extracolsep{\fill}}cccccccccccc@{}}
\hline\hline
{\qquad}& SrSiPd & BaSiPd & CaSiPt & SrSiPt & SrGePt & BaSiPt & BaGePt & ZrSiS\tnote{a} & RhGe\tnote{b} & TaAs\tnote{c} & Pt\tnote{d} \\
\hline
$\sigma_{xy}^z$& 189 & 267 & 171 & 213 & 246 & 183 & 274 & 79 & -139 & -781 & 2139 \\
$\sigma_{xz}^y$& -205 & -196 & -358 & -310 & -365 & -246 & -208 & -280 & 103 & 357 & {\qquad}\\
\hline\hline
\end{tabular*}
 \begin{tablenotes}
        \footnotesize
        \item[a]\emph{Ab initio} calculation~\cite {yen2020tunable};
        \item[b]\emph{Ab initio} calculation~\cite{hsieh2022helicity};
        \item[c]\emph{Ab initio} calculation~\cite{sun2016strong};
        \item[d]\emph{Ab initio} calculation~\cite{guo2008intrinsic}.
      \end{tablenotes}
  \end{threeparttable}
  \label{tI}
\end{table*}

In order to obtain the qualitative insight into the multi-fold degenerate points, we build the low-energy effective model for the states around them~\cite{bradley2010mathematical,yu2022encyclopedia} as shown in Fig.~\ref{fig2}. From a symmetry point of view, the little group at the $\Gamma$ point is $T$ and the double irreducible representation is the combination of R$_5$ and R$_6$. Based on the symmetries, the effective Hamiltonian model can be generated as:
\begin{eqnarray}
H_{RF}&=&a_{1}\Gamma_{0,0}+a_{2}\left(\Gamma_{3,1} q_{x}-\Gamma_{0,2} q_{y}-\Gamma_{3,3} q_{z}\right)+[\alpha_{1}(q_{x}\Gamma_{+, 3} \cr\cr &-&e^{i \pi / 6} q_{y} \Gamma_{+, 0}-e^{i \pi / 3} q_{z} \Gamma_{+, 1})+ h.c. ]
\end{eqnarray}
where the energy and the momentum $q$ are measured from $\Gamma$ point; $\alpha_{1}$ and $a_{i}$$\text {'s }$ represent complex parameter and real parameters, respectively; $\Gamma_{i,j}$=$\sigma_{i}\otimes\sigma_{j}$ with $\sigma_{0}$ being the identity matrix and $\sigma_{i}$ being the pauli matrices. Here, we have $\Gamma_{0,0}$=$\sigma_{0}\otimes\sigma_{0}$, $\Gamma_{+,3}$=$\sigma_{+}\otimes\sigma_{3}$, $\Gamma_{3,3}$=$\sigma_{3}\otimes\sigma_{3}$, $\Gamma_{+,0}$=$\sigma_{+}\otimes\sigma_{0}$ and $\Gamma_{+,1}$=$\sigma_{+}\otimes\sigma_{1}$ with $\sigma_{\pm}=\left(\sigma_{1}\pm i\sigma_{2}\right)/2$. By fitting the band structure from first-principles calculations, the extracted model parameters are $\alpha_{1}$ = 1.89 eV \AA, $a_{1}$ = 1.63 eV \AA, and $a_{2}$ = 1.12 eV \AA.

For the six-fold nodal point at R, it can be regarded as two copies of spin-1 fermions with the irreducible representations of R$_7$. The symmetries for protecting the six-fold degenerate point include ${C_{3}^{111}}$, ${S}_{2x}$, ${S}_{2y}$ and time reversal symmetry $\mathcal{T}$.
Therefore, we can derive the following $k \cdot $p  effective model for the $R$ point:
\begin{align}\nonumber
H_{SF}=&b_{1}S_{0,0}+b_{2}\left(S_{0,1}q_{x}-S_{0,2}q_{y}+S_{0,3} q_{z}\right)\\ +&\sum_{i=1}^{3}c_{i}\left(S_{i, 4}q_{x}+S_{i, 6}q_{y}+S_{i, 7}q_{z}\right)\tag{4}
\end{align}
where the energy and the momentum $q$ are measured from $R$ point; $b_{i}$ and $c_{i}$$\text {'s }$ represent real parameters. $S_{i,j}=\sigma_{i}\otimes$A$_{j}$ is six-dimensional matrixes. The matrix forms of A$_{j}$ are listed in Appendix~\ref{s1}. By fitting the band structure from DFT calculation, the obtained model parameters are $b_{1}$ = 2.99 eV \AA, $b_{2}$ = 12.01 eV \AA, $c_{1}$ = 1.13 eV \AA, $c_{2}$ = 23.16 eV \AA, and $c_{3}$ = 31.33 eV \AA.

In Fig.~\ref{fig3}, we address the results of surface spectra in SrGePt for the (010) surface. Well agreement of bulk band structures has been obtained between first-principles calculation and Wannier-TB model (see Appendix~\ref{wannier-TB}). The calculated surface states along high symmetry directions are plotted in Fig.~\ref{fig3}(a), which demonstrates one of nontrivial surface states at $\bar{\Gamma}-\bar{\rm{Z}}$ originating from four-fold nodal point at $\bar{\Gamma}$. Furthermore, we show isoenergy surface-state spectrum at the energy position of four-fold nodal point as illustrated in Fig.~\ref{fig3}(b), clearly exhibiting four Fermi arcs emanating from the center of surface BZ ($\bar{\Gamma}$), which agrees with the Chern number for spin-3/2 RSW fermion. Fig.~\ref{fig3}(c) plots surface contour at $\rm E_{W_{2}}$, showing that the arcs emerge from the projected Weyl points ($\rm W_{2}$) and hybrid with bulk states, which is similar to the cases in ${\mathrm{WTe}}_{2}$~\cite{soluyanov2015type}, ${\mathrm{MoTe}}_{2}$~\cite{PhysRevB.92.161107}, ${\mathrm{TaIrTe}}_{4}$~\cite{PhysRevB.93.201101} and $X{\mathrm{P}}_{2}$ ($X=\mathrm{Mo}$, W)~\cite{PhysRevLett.117.066402}. It should be noted that the fermi arcs of sixfold excitation at the R point can not distinguished well since they strongly overlap with the trivial states [see Fig.~\ref{fig3} (d)]. Bulk band structure of SrGePt is further checked via the mBJ potential (see Appendix~\ref{s3}), in which the essential topological characteristics remain unchanged.

\section{Spin Hall Effect}

\begin{table*}[t!]
\renewcommand\arraystretch{1.4}
\caption {{Calculated SHC for the SrGePt-family materials as chemical potential is shifted to  $\rm E_{RF} $ and $\rm E_{SF} $. The value of  $\rm E_{RF} $ and $\rm E_{SF} $ are shown in brackets. Here SHC and energy have units of $(\hbar/e)(\Omega \cdot cm)^{-1}$ and eV, respectively.}}
\begin{tabular*}{\hsize}{@{}@{\extracolsep{\fill}}ccccccccccccccc@{}}
        \hline\hline
     {\qquad} & \multicolumn{2}{c}{SrSiPd} & \multicolumn{2}{c}{BaSiPd} & \multicolumn{2}{c}{CaSiPt}& \multicolumn{2}{c}{SrSiPt}& \multicolumn{2}{c}{SrGePt}& \multicolumn{2}{c}{BaSiPt}& \multicolumn{2}{c}{BaGePt}\\
      {\qquad}& $\rm E_{RF}$ & $\rm E_{SF}$&$\rm E_{RF}$ & $\rm E_{SF}$& $\rm E_{RF}$ & $\rm E_{SF}$ & $\rm E_{RF}$ & $\rm E_{SF}$ & $\rm E_{RF}$ & $\rm E_{SF}$ & $\rm E_{RF}$ & $\rm E_{SF}$ & $\rm E_{RF}$ & $\rm E_{SF}$\\
        {\qquad} & (-0.372) & (0.470)& (-0.228) & (0.275)  & (-0.488) & (0.416) & (-0.365) & (0.338)  & (-0.393) & (0.402)  & (-0.159) & (0.203)& (-0.207) & (0.246)\\
     \hline
      $\sigma_{xy}^z$ & 332 & 73 & 257 & 13 & 152 & -91 & 145 & 126 & 185  & 16 & 204 & 152 & 256 & 226  \\
      $\sigma_{xz}^y$ & -572 & -127 & -358 & -106 & -477 & -343 & -271 & -260 & -481  & -397 & -295 & -156 & -270 & -187 \\
       \hline\hline
       \end{tabular*}
       \label{tII}
\end{table*}

The intrinsic SHC is a significant physical quantity to characterize the strength of the intrinsic SHE in a material. SHC belongs to third-order tensor, which generally contains 27 elements. Due to the constraint of crystal symmetries in SrGePt~\cite{seemann2015symmetry}, it reduces to only two independent nonzero components, i.e. $\sigma_{xy}^z$ and $\sigma_{xz}^y$.
$\sigma_{xy}^z$ and $\sigma_{xz}^y$ at the Fermi level have been calculated for all seven compounds of the SrGePt family materials in increasing ordering of SOC strength. They are listed in Table~\ref{tI} along with SHC of ZrSiS, RhGe, TaAs and Pt for comparisons. We find that CaSiPt hosts the smallest magnitude of $\sigma_{xy}^z$ and BaGePt has the biggest one, exhibiting the variation trend with the increasing order of SOC strength. Nevertheless, the size of $\sigma_{xz}^y$ varies with the strength of SOC in a complex way. Besides, one can observe that the amplitude of the calculated $\sigma_{xy}^z$ for each member of the SrGePt-family is larger than those of ZrSiS and RhGe and much smaller than those of TaAs and Pt, whereas $\sigma_{xz}^y$ is comparable to those of ZrSiS, RhGe and TaAs. The signs of $\sigma_{xy}^z$ and $\sigma_{xz}^y$ are positive and negative for the SrGePt-family materials, respectively.

Furthermore, $\sigma_{xy}^z$ and $\sigma_{xz}^y$ of the SrGePt-family materials are listed in Table~\ref{tII} as the chemical potential ($\mu$) is shifted to the  $\rm E_{RF} $ and $\rm E_{SF} $. Also, their band structures are shown in Appendix~\ref{wsoc}. One can see that the magnitude of $\sigma_{xy}^z$ and $\sigma_{xz}^y$ of the most SrGePt-family materials at $\mu$=$\rm E_{F} $ is larger than those at $\mu$=$\rm E_{SF} $, while they are smaller than those at $\mu$=$\rm E_{RF} $. Such varying behaviors suggest that SHC can be effectively modulated by tuning the chemical potential. More remarkably, the results as shown in Fig.~\ref{shc} and Fig.~\ref{all_SHC} of Appendix~\ref{six_shc} indicate that the magnitudes of $\sigma_{xy}^z$ and $\sigma_{xz}^y$ for SrSiPd, BaSiPd and BaSiPt decrease steeply as $\mu$ is raised from $\rm E_{RF} $ to $\rm E_{SF} $, while those for the other four compounds exhibit relatively flat trend. Moreover, the magnitudes of $\sigma_{xy}^z$ and $\sigma_{xz}^y$ for the SrGePt-family materials vary dramatically around $\mu$=$\rm E_{RF} $, which could be the common features of SHC for spin-3/2 RSW fermions as reported in Refs.~\cite{PhysRevResearch.3.033101,hsieh2022helicity}.

\begin{figure}[htp!]
	{\includegraphics[clip,width=8.2cm]{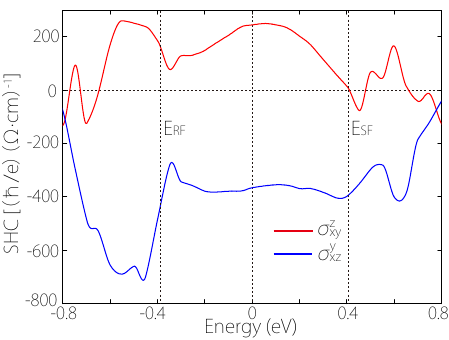}}
	\caption{\label{shc} $\sigma_{xy}^{z}$ and $\sigma_{xz}^{y}$ as a function of chemical potential for SrGePt.}
\end{figure}

\begin{figure}[htp!]
	{\includegraphics[clip,width=8.2cm]{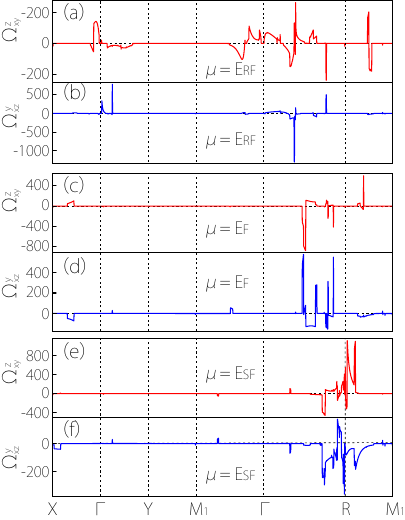}}
	\caption{\label{omega} (a) The calculated $\Omega_{xy}^{z}$ [$\Omega_{xz}^{y}$] at (a) [(b)] $\mu$=$\rm E_{RF} $, (c) [(d)] $\mu$=$\rm E_{F} $ and (e) [(f)] $\mu$=$\rm E_{SF} $ along the high-symmetry directions for SrGePt.}
\end{figure}

\begin{figure}[htp!]
	{\includegraphics[clip,width=8.2cm]{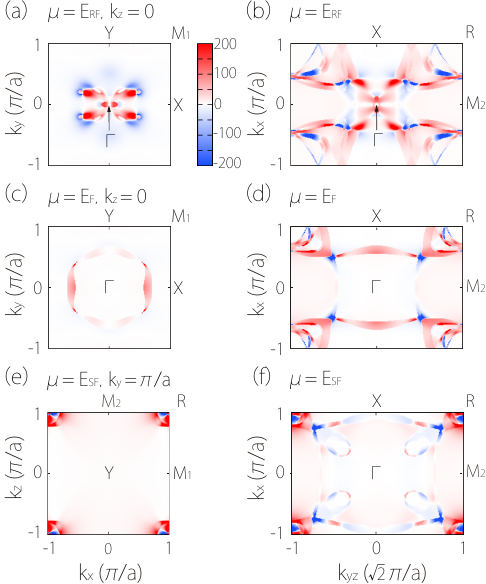}}
	\caption{\label{sbcxyz} The distributions of the calculated $\Omega_{xy}^{z}$ for 2D slices of (a) [(c)] $k_z$=0 and (b) [(d)] $k_x$-$k_{yz}$ at $\mu$=$\rm E_{RF} $ [$\rm E_{F} $], as well as 2D slices of (e) $k_y$=$\pi/a$ and (f) $k_x$-$k_{yz}$ at $\mu$=$\rm E_{SF} $ for SrGePt. Here the direction of $k_{yz}$ is along (011)-axis.}
\end{figure}

\begin{figure}[htp!]
	{\includegraphics[clip,width=8.2cm]{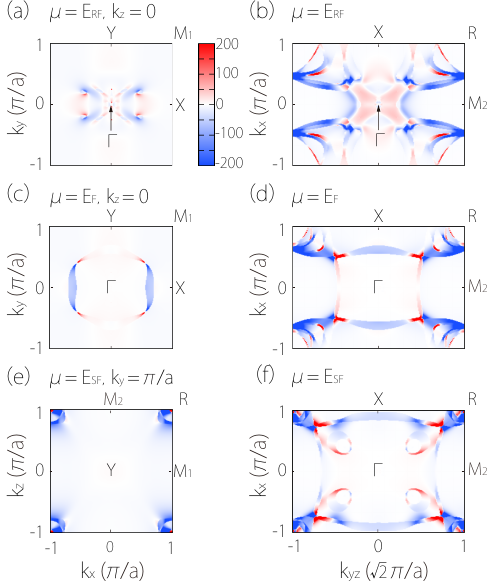}}
	\caption{\label{sbcxzy} The distributions of the calculated $\Omega_{xz}^{y}$ for 2D slices of (a) [(c)] $k_z$=0 and (b) [(d)] $k_x$-$k_{yz}$ at $\mu$=$\rm E_{RF} $ [$\rm E_{F} $], as well as 2D slices of (e) $k_y$=$\pi/a$ and (f) $k_x$-$k_{yz}$ at $\mu$=$\rm E_{SF} $ for SrGePt. Here the direction of $k_{yz}$ is along (011)-axis.}
\end{figure}

Next, we take SrGePt as an example to elucidate the impact of the electronic states around spin-3/2 RSW fermion and sixfold excitation on SHC. In Fig.~\ref{shc}, we plot the results for the SHC of SrGePt as a function of chemical potential. One can observe that the value of $\sigma_{xz}^y$ almost remains unchanged, while that of $\sigma_{xy}^z$ decrease for raising the chemical potential to $\rm E_{SF} $ or lowering it to $\rm E_{RF} $. For instance, the value of $\sigma_{xz}^y$ can be tuned from -365 $(\hbar/e)(\Omega \cdot cm)^{-1}$ to -481 $(\hbar/e)(\Omega \cdot cm)^{-1}$ for varying the chemical potential from $\rm E_{F} $ to $\rm E_{RF} $, which can be realized via hole doping of about 0.31 e/f.u. As $\mu$=$\rm E_{F} $, we find that the magnitude of $\sigma_{xz}^y$ (365 $(\hbar/e)(\Omega \cdot cm)^{-1}$) is larger than that of $\sigma_{xy}^z$  (246 $(\hbar/e)(\Omega \cdot cm)^{-1}$), indicating the anisotropic characteristic of spin transport properties.

To elucidate the underlying mechanism of the large SHC in SrGePt, we evaluate the $\mathbf{k}$-resolved $\Omega_{xy}^{z}(\mathbf{k})$ and $\Omega_{xz}^{y}(\mathbf{k})$ along the high symmetry lines at $\rm E_{RF} $, $\rm E_{F} $ and $\rm E_{SF} $ as depicted in Fig.~\ref{omega}. It is clear to see that both $\Omega_{xy}^{z}(\mathbf{k})$ and $\Omega_{xz}^{y}(\mathbf{k})$ strongly depend on wave vector $\mathbf{k}$, showing prominent peaks around nodal-points, which mainly contribute to the intrinsic SHC. It should be noted that SBC exhibits the strong anisotropy along distinct directions such as $\Gamma$-X and $\Gamma$-Y, owing to the absence of four-fold symmetry.
Furthermore, we show the $\mathbf{k}$-resolved SBC $\Omega_{xy}^z$ and $\Omega_{xz}^y$ for different 2D BZ planes as $\mu$=$\rm E_{RF} $, $\mu$=$\rm E_{F}$ and $\mu$=$\rm E_{SF} $ (see Figs.~\ref{sbcxyz} and ~\ref{sbcxzy}). One can observe that the distributions of SBC $\Omega_{xy}^z$ ($\Omega_{xz}^y$) are mainly contributed by the positive (negative) value originating from the SOC-induced band-splitting and nodal-points, which eventually leads to the positive (negative) SHC. Specially, the large peaks appear around the $\Gamma$ point at $\mu$=$\rm E_{RF} $, which relevant with the presence of four-fold nodal-point. While $\mu$=$\rm E_{SF} $, the prominent peaks occurred around the R point, which is associated with appearance of six-fold nodal-point. Moreover, multiple peaks always happened along the $\Gamma$-R direction due to the Weyl points. These large SBC peaks can greatly impact on the amplitude of SHC. It is worthwhile to mention that the value of SHC almost retains unchanged for the optimized geometry and experimental geometry (see Appendix~\ref{opt_shc}). The large and tunable SHC as well as the strong SOC in the SrGePt family materials provide promising applications for employing the chiral topological materials in spintronics.



\begin{table*}[t!]
	\renewcommand\arraystretch{1.4}
	\caption {{The experimental ($\rm \mathnormal a_{exp}$) and simulated ($\rm \mathnormal a_{sim}$) lattice constants in the SrGePt family.}}
	\begin{tabular*}{\hsize}{@{}@{\extracolsep{\fill}}cccccccc@{}}
		\hline\hline
		Materials &SrSiPd& BaSiPd& CaSiPt& SrSiPt & SrGePt& BaSiPt& BaGePt\\
		\hline
		$\rm \mathnormal a_{exp} \left[\AA\right]$& 6.500& 6.662& 6.320& 6.485& 6.602& 6.633& 6.747\\
		$\rm \mathnormal a_{sim} \left[\AA\right]$& 6.570& 6.750& 6.385& 6.553& 6.692& 6.717& 6.851\\
		\hline\hline
	\end{tabular*}
\label{tIII}
\end{table*}

\section{Conclusions}

In conclusions, based on first-principles calculations we have predicated topological metals with multiple types of quasiparticle fermions in the SrGePt family materials, which have been synthesized experimentally. We found that electronic structures in SrGePt show the coexistence of fourfold spin-3/2 RSW fermion, sixfold excitation, and Weyl fermions, which are guaranteed by nonsymorphic and broken-inversion symmetries. Long Fermi arcs originating from the spin-3/2 RSW fermion are clear to appear on the (010) surface at the energy of such fermions, which span the whole BZ. Additionally, Fermi arcs from the projected Weyl points W$_2$ are shown, which overlap with bulk states. Furthermore, we revealed that they host the remarkable SHC evaluated via Kubo formula, which originates from the large SBC around the nodal points and the strong SOC-induced band-splitting. We expect that our work can simulate the experiments on the new chiral topological metals for the surface spectra and the large SHE.

\appendix
\section{Experimental and simulated constants of materials}

In Table~\ref{tIII}, we list simulated lattice constants with optimized unit cell and atomic positions for the SrGePt-family materials, which are compared with experimental ones. In this work, we adopt the simulated lattice constants for all the first-principles calculations.
\label{simu_con}

\begin{figure}[htp!]
	{\includegraphics[clip,width=8.2cm]{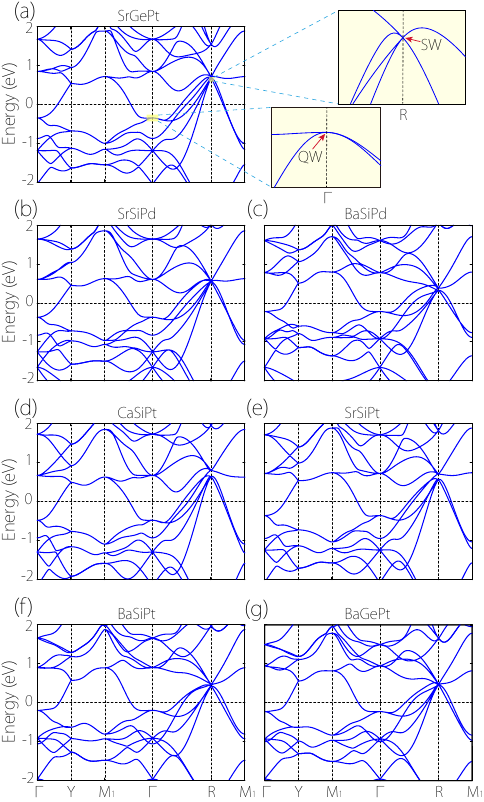}}
	\caption{\label{band_wosoc} Bulk band structures without SOC for (a) SrGePt, (b) SrSiPd, (c) BaSiPd, (d) CaSiPt, (e) SrSiPt, (f) BaSiPt and (g) BaGePt. Two insets indicate the enlargement of bands around SW at R and QW at $\Gamma$, respectively.}
\end{figure}

\section{Band structures without SOC for the SrGePt-class materials}

The calculated band structures of all the SrGePt-family materials without SOC included are shown in Fig.~\ref{band_wosoc}. The main features of band structures include quadruple Weyl node with $\mathcal{C}$=4 (QW) at the $\Gamma$ point and spin-1 Weyl node with $\mathcal{C}$=2 (SW) at the R point~\cite{PhysRevB.102.125148}, as show in zoom-in of Fig.~\ref{band_wosoc}. Once SOC is turned on, QW will transform into spin-3/2 RSW nodal point and SW will split into six-fold nodal point and Weyl point, respectively.
\label{wosoc}

\section{Distributions of Berry curvature around Weyl points W$_2$ and W$_3$}

\begin{figure}[htp!]
	{\includegraphics[clip,width=8.2cm]{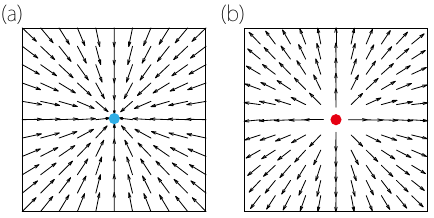}}
	\caption{\label{BC} Distributions of Berry curvature around Weyl points (a) W$_2$ and (b) W$_3$.}
\end{figure}
Fig.~\ref{BC} shows the distributions of Berry curvature around Weyl points W$_2$ and W$_3$, which act as the sink and source in the \emph{k}-space. These results are consistent with the Chern number of -1 and +1 for W$_2$ and W$_3$, respectively.
\label{BC_WP}

\section{Band structures for other members of the SrGePt-class materials}
\begin{figure}[htp!]
	{\includegraphics[clip,width=8.2cm]{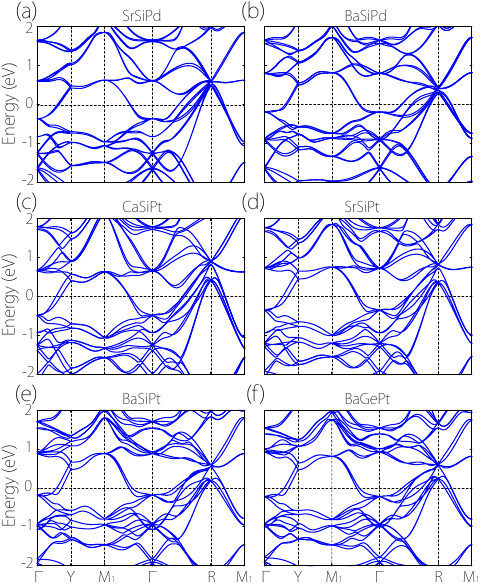}}
	\caption{\label{fig5} Bulk band structures with SOC included for (a) SrSiPd, (b) BaSiPd, (c) CaSiPt, (d) SrSiPt, (e) BaSiPt and (f) BaGePt.}
\end{figure}
\label{wsoc}
We have investigated the bulk band structures with SOC included for other members of the SrGePt family materials, as shown in Fig.~\ref{fig5}. One can see that all of them feature (unconventional) quasiparticle fermions including fourfold spin-3/2 RSW fermion, sixfold excitation, and Weyl fermions in the vicinity of Fermi level, similar to those in SrGePt. The corresponding nontrivial surface states like the case of SrGePt are expected to appear on the side surface.

\section{Matrix form for A$_{j}$}
Here $A_{0}$ is the 3$\times$3 identity matrix, and the other 3$\times$3 matrices take the following forms:

\begin{equation}
\begin{aligned}\nonumber
A_1=&\begin{bmatrix}
0 & -i & 0 \\
i & 0 & 0 \\
0 & 0 & 0
\end{bmatrix}~\\\nonumber
\end{aligned},
\begin{aligned}\nonumber
A_2=&\begin{bmatrix}
0 & 0 & -i \\
0 & 0 & 0 \\
i & 0 & 0
\end{bmatrix}~\\\nonumber
\end{aligned},
\end{equation}
\begin{equation}
\begin{aligned}\nonumber
A_3=&\begin{bmatrix}
0 & 0 & 0 \\
0 & 0 & -i \\
0 & i & 0
\end{bmatrix}~\\\nonumber
\end{aligned},
\begin{aligned}\nonumber
A_4=&\begin{bmatrix}
0 & 1 & 0 \\
1 & 0 & 0 \\
0 & 0 & 0
\end{bmatrix}~\\\nonumber
\end{aligned},
\end{equation}
\begin{equation}
\begin{aligned}\nonumber
A_5=&\begin{bmatrix}
1 & 0 & 0 \\
0 & -1 & 0 \\
0 & 0 & 0
\end{bmatrix}~\\\nonumber
\end{aligned},
\begin{aligned}\nonumber
A_6=&\begin{bmatrix}
0 & 0 & 1 \\
0 & 0 & 0 \\
1 & 0 & 0
\end{bmatrix}~\\\nonumber
\end{aligned},
\end{equation}
\begin{equation}
\begin{aligned}\nonumber
A_7=&\begin{bmatrix}
0 & 0 & 0 \\
0 & 0 & 1 \\
0 & 1 & 0
\end{bmatrix}~\\\nonumber
\end{aligned},
\begin{aligned}\nonumber
A_8=\frac{1}{\sqrt{3}}&\begin{bmatrix}
1 & 0 & 0 \\
0 & 1 & 0 \\
0 & 0 & -2
\end{bmatrix}~\\\nonumber
\end{aligned}.
\end{equation}
\label{s1}


\section{Band structures of the SrGePt-class materials calculated by first-principles and Wannier-TB}
\begin{figure}[htp!]
	{\includegraphics[clip,width=8.2cm]{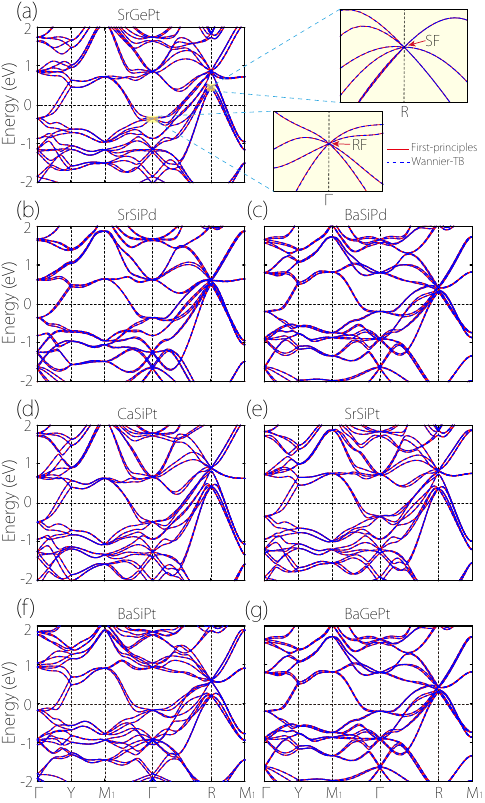}}
	\caption{ Comparisons of calculated band structures from first-principles and Wannier-TB for (a) SrGePt, (b) SrSiPd, (c) BaSiPd, (d) CaSiPt, (e) SrSiPt, (f) BaSiPt and (g) BaGePt. Two insets show the enlarged view of two yellow boxes. }
	\label{w-TB}
\end{figure}

Fig.~\ref{w-TB} shows the results of calculated band structures from Wannier-TB and first-principles for all seven compounds of the SrGePt-family materials. Well agreements are obtained between them.
\label{wannier-TB}

\section{Band structure of SrGePt with mBJ potential}
\begin{figure}[htp!]
	{\includegraphics[clip,width=6.2cm]{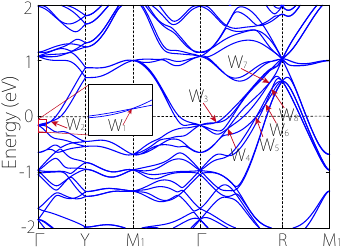}}
	\caption{\label{fig4}Band structure of SrGePt calculated by mBJ potential.}
\end{figure}
\label{s3}

\begin{figure}[b!]
	{\includegraphics[clip,width=8.2cm]{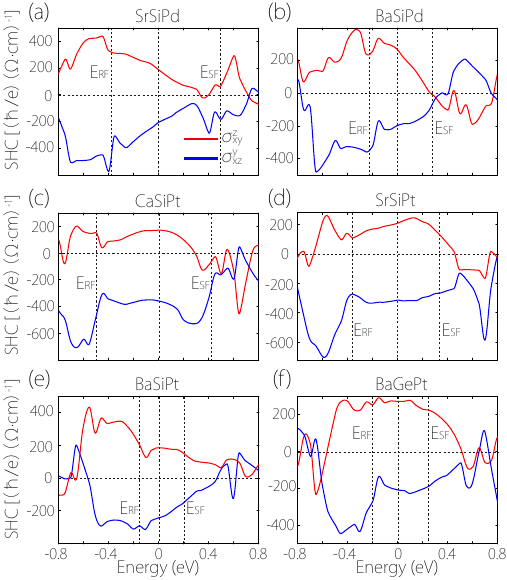}}
	\caption{\label{all_SHC}Calculated $\sigma_{xy}^z$ (red line) and $\sigma_{xz}^y$ (blue line) as a function of chemical potential for (a) SrSiPd, (b) BaSiPd, (c) CaSiPt, (d) SrSiPt, (e) BaSiPt and (f) BaGePt.}
\end{figure}
In order to check the validity of our results, we further calculated the bulk band structure for SrGePt through the mBJ potential, as shown in Fig.~\ref{fig4}. It is clear to see that the band structure is totally agreement with that using PBE potential (see ~Fig.~\ref{fig2}), and the main features of the crossing points are still maintained.

\section{SHC for other six compounds of the SrGePt-family materials}

In Fig.~\ref{all_SHC}, we plot the results of SHC as a function of chemical potential for other six compounds of the SrGePt-family materials. It is clear to show that the magnitudes of $\sigma_{xy}^z$ and $\sigma_{xz}^y$ for SrSiPd, BaSiPd and BaSiPt decrease steeply as $\mu$ is raised from $\rm E_{RF}$ to $\rm E_{SF}$, while those for the other three compounds exhibit relatively flat trend.
\label{six_shc}

\section{Comparisons of SHC between simulated geometry and experimental geometry for SrGePt}
\begin{figure}[htp!]
	{\includegraphics[clip,width=6.2cm]{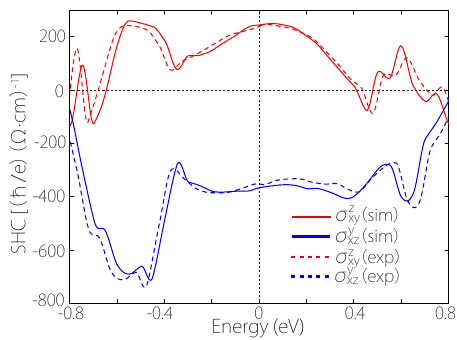}}
	\caption{\label{exp_opt}Calculated SHC of SrGePt materials for simulated geometry (solid lines) and experimental geometry (dotted lines).}
\end{figure}
The results of calculated SHC with optimized structure and experimental structure in SrGePt are depicted in Fig.~\ref{exp_opt}. One can see that the value of SHC almost retains unchanged after the structure of SrGePt is fully relaxed.
\label{opt_shc}

\begin{acknowledgements}
	The authors thank Shengyuan A. Yang for the valuable discussions. This work is supported by the Project of Educational Commission of Hunan Province of China (Grant No. 21A0066), the Hunan Provincial Natural Science Foundation of China (Grant No. 2022JJ30370) and the National Natural Science Foundation of China (Grants No. 12174098 and No. U19A2090). We acknowledge computational support from H2 clusters in Xi'an Jiaotong University.
\end{acknowledgements}

\bibliography{chiral_refs}

\bibliographystyle{apsrev4-1}

\end{document}